\documentstyle[preprint,aps]{revtex}
\begin{document}
\draft
\title{A vanishing cosmological constant in elementary particle theory}
\author{F. Pisano} 
\address{Instituto de F\'\i sica Te\'orica,  
Universidade Estadual Paulista \\  
Rua Pamplona 145 -- 01405-000, S\~ao Paulo, SP, Brazil~\footnote{Address 
after March 1997: {\it Instituto de F\'\i sica, Universidade Federal do 
Paran\'a, 81531-990, Curitiba, PR, Brazil}}} 
\author{M.D. Tonasse}
\address{Instituto de F\'\i sica, Universidade do Estado do Rio de Janeiro, \\ 
Rua S\~ao Francisco Xavier 524 --  20550-013, Rio de Janeiro, RJ, Brazil}
\maketitle
\begin{abstract}
The quest of a vanishing cosmological constant is considered in the 
simplest anomaly-free chiral gauge extension of the electroweak 
standard model where the new physics is limited to a well defined 
additional flavordynamics above the Fermi scale, namely up to a few TeVs 
by matching the gauge coupling constants at the electroweak scale, 
and with an extended scalarland. In   
contrast to the electroweak standard model, it is shown how the extended 
scalar sector of the theory allows a vanishing or a very small 
cosmological constant. The details of the cancellation mechanism are 
presented. At accessible energies the theory is 
indistinguishable from the standard model of elementary particles and it is 
in agreement with all existing data.
\end{abstract}
\bigskip
All astronomical surveys agree that there is no evidence for any spacetime 
distortion due to a nonvanishing cosmological constant~\cite{Loh86} 
which is many orders of magnitude smaller than that estimated in theories 
of elementary particles. Up to distances which are 
accessible to astronomers, about  
10 billion light-years, or $10^{28}$ cm, the magnitude of the cosmological 
constant must be smaller than $10^{-56}\,\mbox{cm}^{-2}\approx 10^{-84} 
\,\mbox{GeV}^2$. The possible presence of the cosmological constant $\Lambda$ 
in the Einstein's field equations 
\begin{equation}  
R_{\mu\nu} - \left (\frac{1}{2}R - \Lambda \right ) g_{\mu\nu} = 
\frac{8\pi\,G}{c^4} T_{\mu\nu}
\label{um}
\end{equation}
can eventually be vindicated by measuring exponential deviations from the 
standard matter dominated spatially flat Friedman-Robertson-Walker universe 
scale factor $R(t)\sim t^{2/3}$. The influence of matter on the metric is 
determined by the energy-momentum tensor $T_{\mu\nu}$. The component 
$T_{00}$ is the energy density and the coefficient $c^4/8\pi\,G \approx 
5\times 10^{45}\,\mbox{g}\times\mbox{cm}\times\mbox{sec}^{-2}$, where $G$ 
is the Newtonian gravitational constant, measures the elasticity of the 
vacuum. The cosmological constant is, according to astronomical evidence, 
very close to zero. There is no understanding of why $\Lambda$ should be  
close or equal to zero. This question of the cosmological constant problem, 
which consists in understanding a possible cancellation, is widely regarded as 
one of the most significative mysteries of the modern cosmology. A proposal  
for solution, due to Baum-Hawking-Coleman~\cite{Baum}, has attracted much 
attention. It argues that the vanishing of the cosmological constant is   
closely related to a wormhole-induced quantum instability of the theory.  
The observable value of the cosmological constant could not be an absolutely   
fundamental c-number parameter but a dynamical quantum variable with the 
meaning of topological changes, such as baby universes and 
wormholes~\cite{Hawking87}.    
Nowhere, in physics, we find a greater divergence between theory and 
experiment than in the cosmological constant problem. As emphasized by  
Weinberg~\cite{Weinberg96}, for a particle physicist, all the values in the 
observationally allowed range, extending up to values that would make up 
most of the critical density required in a spatially flat Robertson-Walker 
universe, seem ridiculously implausible. We should remark that all the  
standard cosmological model is faced by a severe hierarchical problem, 
the vacuum energy of the actual universe is extremely fine tuned to 
zero~\cite{Zeldovich68}. 
\par
In elementary particle theory the underlying gauge symmetry is larger than 
that of the actual vacuum whose symmetry is the combination of the color 
and the abelian electromagnetic factors 
$G_{3_C1_{\rm em}}\equiv \mbox{SU(3)}_C\otimes \mbox{U(1)}_{\rm em}$. 
The full gauge symmetry of the standard model for the nongravitational 
interactions~\cite{Glashow61} 
$$
G_{3_C2_L1_Y}\equiv \mbox{SU(3)}_C\otimes \mbox{SU(2)}_L \otimes \mbox{U(1)}_Y
$$
is restored at the Fermi scale 
$$
\Lambda_F = \frac{1}{2^{1/4}\sqrt{G_F}} \approx 246\,\mbox{GeV}
$$
($G_F\approx 1.166\times 10^{-5}\,\mbox{GeV}^{-2}$) equivalent to a 
time of order 
$10^{-11}$ sec. This scale is set by the decay constant of the three 
Goldstone bosons transformed via the Higgs-Kibble mechanism~\cite{Higgs64} 
into the longitudinal components of the weak gauge bosons. The underlying 
$G_{3_C2_L1_Y}$ symmetry is not manifest in the 
structure of the vacuum and nature 
realizes the mechanism of spontaneous symmetry breaking 
$G_{3_C2_L1_Y}\rightarrow G_{3_C1_{\rm em}}$ where the 
$G_{3_C2_L1_Y}$ symmetry is broken because 
the associated vacuum state is not invariant anymore. Spontaneous symmetry 
breaking preserves the renormalizability of the original gauge theory even 
after symmetry breaking, giving us a renormalizable theory of massive 
vector bosons~\cite{tHooft71}. 
\par
Let us consider a scalar field with a $\phi^4$ interaction 
\begin{eqnarray}
{\cal L} & = & \partial_\mu\phi\partial^\mu\phi - V(\phi); \\ \nonumber 
V(\phi) & = & -\frac{1}{2}m^2\phi^2 + \frac{1}{4}\lambda\,\phi^4
\label{dois}
\end{eqnarray}
containing the discrete symmetry $\phi\rightarrow -\phi$, 
$$
{\cal L}(\phi)\rightarrow {\cal L}(-\phi) = {\cal L}(\phi). 
$$
In the electroweak standard model with the multiplet of scalar fields 
transforming under $G_{3_C2_L1_Y}$ as 
\begin{equation}
\Phi \equiv 
\left (
\begin{array}{c}
\phi^+ \\
\phi^0
\end{array}
\right ) \sim({\bf 1},{\bf 2},+1)
\label{tres}
\end{equation}
the scalar potential can be written in a $\phi^4$ fashion 
\begin{equation}
V(\Phi^\dagger\Phi) = a \Phi^\dagger\Phi + b(\Phi^\dagger\Phi)^2.   
\label{quatro}
\end{equation}
After the process of spontaneous symmetry breaking, the neutral component of 
the scalar doublet gets a vacuum expectation value so that 
\begin{equation}
\langle \Phi \rangle = \langle\phi^0\rangle = \frac{1}{\sqrt 2} 
\left ( 
\begin{array}{c}
0 \\
\sigma
\end{array}
\right )
\label{cinco}
\end{equation}
and in terms of $\sigma$ the potential is 
$$
V(\sigma) = \frac{a}{2}\sigma^2 + \frac{b}{4}\sigma^4. 
$$
Now we define 
$$
a \equiv - m^2, \quad b \equiv \lambda;\quad m > 0
$$  
and the potential becomes 
\begin{equation}
V(\sigma) = -\frac{1}{2}m^2\sigma^2 + \frac{1}{4}\lambda\sigma^4 
\label{seis}
\end{equation}
with the minima determined by the conditions 
$$
V^\prime\equiv \frac{\partial V}{\partial\sigma} = 
\sigma (-m^2 + \lambda\sigma^2) = 0
$$
and 
$$
V^{\prime\prime} \equiv \frac{\partial^2 V}{\partial\sigma^2} = 
-m^2 + 3\lambda\sigma^2 > 0
$$
which show that a particle would rather not sit at the vacuum state 
$\phi = 0$. Instead, it moves down the potential to a lower-energy state 
given by the botton of one of the wells 
\begin{equation}
\sigma_\pm = \pm \left (\frac{m^2}{\lambda}\right )^\frac{1}{2}
\label{sete}
\end{equation}
where the potential takes the value 
\begin{equation}
V(\sigma_\pm) = - \frac{m^4}{4\lambda}
\label{oito}
\end{equation}
and $V^{\prime\prime} (\sigma_\pm) = 2m^2$ is 
the curvature of the potential about the true ground state associated 
to the mass $M$ of the physical boson by 
\begin{equation} 
M^2 = V^{\prime\prime}(\sigma_\pm) = 2 m^2 = 2\lambda \sigma_\pm^2. 
\label{nove}
\end{equation}
Further, we make the simple choice   
\begin{equation}
\Phi = \frac{1}{\sqrt 2} 
\left (
\begin{array}{c}
0 \\
\sigma + H
\end{array}
\right )
\label{dez}
\end{equation}
where $H$ is the neutral physical Higgs boson. The potential given in 
Eq.~(\ref{quatro}) would become at tree level 
\begin{equation}
V(H) = - \frac{m^4}{4\lambda} - m^2\,H^2 + \lambda\sigma\,H^3 + 
\frac{\lambda}{4}\,H^4
\label{onze}
\end{equation}
then, with Eq.~(\ref{nove}), we get 
\begin{equation}
V(H) = - \frac{m^4}{4\lambda} - \frac{1}{2}\,M^2\,H^2 + \lambda\sigma\,H^3 + 
\frac{\lambda}{4}\,H^4. 
\label{doze}
\end{equation}
The field independent constant term $-m^4/4\lambda$ preceding the mass term is 
the energy density of the vacuum given by the component 
\begin{equation}
\langle T^0_{\,0}\rangle \equiv \rho_\Lambda 
\propto V(\sigma_\pm) = -\frac{m^4}{4\lambda}
\label{treze}
\end{equation}
of the stress tensor for a scalar field. Following 
Zel'dovich~\cite{Zeldovich68}, the contribution of the vacuum energy density 
plays the role of a cosmological constant 
$$
\Lambda = \frac{8\pi\,G}{c^2}\rho_\Lambda = 
\frac{8\pi\,G}{c^4} V(\sigma_\pm) = 
- \frac{2\pi\,G}{c^4}m^2\sigma^2_\pm
$$
which through the relation involving the Fermi constant and $\sigma$ 
$$
\frac{G_F}{\sqrt 2} = \frac{1}{2\sigma^2}
$$
and Eq.~(\ref{nove}) yields   
\begin{equation}
\Lambda = - \frac{\pi\,G}{2\sqrt 2c^4G_F}\,M^2 \approx -1.3\times 10^{-33}\,M^2
\label{catorze}
\end{equation}
with $\Lambda$ as a functional dependence of $M^2$ only. According to the 
astronomical bound~\cite{Loh86} 
$$
|\Lambda_{\rm obs}| < 10^{-56}\,\mbox{cm}^{-2} 
\approx 10^{-84}\,\mbox{GeV}^2,  
$$
corresponding to 
$$
|\rho_\Lambda| < 5\times 10^{-29}\,\mbox{g}/\mbox{cm}^3 \approx 10^{-47}\,
\mbox{GeV}^4
$$
if $\sigma_\pm$ are absolute minima of the potential, $V(\sigma_\pm) < V(0)$, 
and taking the bound on the Higgs boson mass $M > 10$ GeV~\cite{Kane93} it 
follows that 
\begin{equation}
|\Lambda/\Lambda_{\rm obs}| \approx 10^{52}.
\label{quinze}
\end{equation}
Therefore in the standard model of elementary particles it is expected a 
large vacuum energy. The spontaneous symmetry breaking of the gauge symmetry,      
necessary for generating masses, introduces a vacuum energy into the theory.   
It is tempting to require $\rho_\Lambda = 0$ 
which can be 
accomplished by adding the constant $+m^4/4\lambda$ to the Lagrangian which 
does not affect the equations of motion or the quantization of the theory. 
Spontaneous symmetry breaking implies a large cosmological constant in the 
standard Higgs model. Anyway the supersymmetric standard model as well as 
superstring theories~\cite{Lopez94}               
have this same problem which is the most drastic of all   
fine-tuning and naturalness problem. A mechanism by which some form of 
supersymmetry makes $\rho_\Lambda$ vanish was proposed recently by 
Witten~\cite{Witten95}. Also technicolor with extended technicolor, which 
has spontaneous breaking at least up to 100s of TeV, has a terrible 
cosmological constant problem~\cite{Lane94}. 
\par
There are no experimental data that directly run against the predictions   
of the standard theory of elementary particles. All experimental data are 
consistent with the standard model and so theoretical extensions must be 
motivated by attempting to understand features that are accomodated in the 
standard model but not explained by it. In this respect, many theorists 
believe that there should be some new physics lurking at the TeV scale, 
accessible to the next generation of colliders, or better yet, at present 
colliders such as the Tevatron at Fermilab or LEP II at CERN. The simplest 
chiral gauge extension of the standard model gauge group that one could 
consider is
$$
G_{3_C3_L1_X}\equiv \mbox{SU(3)}_C \otimes \mbox{SU(3)}_L \otimes 
\mbox{U(1)}_X.
$$
Although there exist several models with the $G_{3_C3_L1_X}$ gauge 
symmetry~\cite{Valle80} recent proposals have a different representation 
content and a quite different new physics at no high energy 
scale~\cite{Pisano92}. There exist several distinct possibilities depending 
on the way the electric operator for fermions 
\begin{equation}
{\cal Q}/e = \frac{1}{2}(\lambda_3^L + \xi\lambda_8^L) + X
\label{dezeseis}
\end{equation}
is embedded in the neutral generators of the $G_{3_C3_L1_X}$ group where 
$\xi$ is the embedding parameter. Notice that the electric charge operator 
can be written as
$$
{\cal Q}/e = \frac{1}{2}\lambda_3^L + \frac{1}{2}Y
$$
where
$$
Y = \xi\lambda_8^L + 2X
$$
is the hypercharge operator of the gauge group of the standard model. 
\par
The salient features of these models which clarify loose ends of the standard 
model can be quoted as: 1) In the standard model, each family of fermions is 
anomaly free. This is true for many extensions including grand unified 
theories and supersymmetric models. In the $G_{3_C3_L1_X}$ each family is 
anomalous but different families are not replicas of one another, and the 
anomalies cancel when the number of families are taken into account, and 
to be a multiple of the number of colors. This novel method of anomaly  
cancellation requires that at least one family transforms differently from 
the others, thus breaking generation universality. The $G_{3_C3_L1_X}$ is 
the most economical gauge group which admits such fermion representation 
and gives the first step for understanding the flavor  
question~\cite{Pisano92,Pisano96}; 
2) The $G_{3_C3_L1_X}$ allows to give 
a natural answer to the family replication question and furthermore gives 
some indication as to why the {\it top} quark is so 
heavy~\cite{Pisano92,Pisano94}; 
3) The electroweak mixing angle free parameter, $\theta_W$, of the 
standard model is limited from above. In the $\xi = -\sqrt 3$ and 
$\xi = 1/\sqrt 3$ cases we obtain the bounds $\sin^2\theta_W < 1/4$ and 
$\sin^2\theta_W < 3/4$, respectively~\cite{Montero93}; 4) Electric charge 
quantization is another achievement within $G_{3_C3_L1_X}$ symmetry. In 
particular in models with $0,\,\pm 1$ charged leptons there is always a 
family transforming as $({\bf 1},{\bf 3}_L,0)$ under $G_{3_C3_L1_X}$. In  
this generation there is charge quantization in the fashion of grand 
unified theories. From anomaly cancellation follows the electric charge 
quantization in the other families~\cite{Pisano96}; 5) One of the quark 
families transforms differently from the other two. Using experimental 
input on neutral bosons mixing, the third family must be the one that 
is singled out, at least up to small family mixing~\cite{Liu94}.
\par
The phenomenology of models with extended gauge symmetries provides the 
existence of extra gauge bosons and new exotic fermions and their discovery 
would be a definitive signal of new physics~\cite{Hewett}. Let us 
summarize the principal consequences of $G_{3_C3_L1_X}$ models concerning 
physics beyond the standard model: 1) There are five additional gauge bosons.  
In the model with the embedding parameter $\xi = -\sqrt 3$, there are a 
neutral $Z^\prime$ and four bileptons, $(U^{--}, V^-)$ with lepton number 
$L=+2$ and $(U^{++},V^+)$ with lepton number $L=-2$. Here 
$L=L_e+L_\mu+L_\tau$ is the total lepton number and there is not conservation 
of the family lepton number $L_i$, $(i = e,\,\mu,\,\tau)$~\cite{bilepton}. 
At $e^-p$ colliders such as HERA or LEPII-LHC were considered the prospects 
of searching for bileptons. At LEPII-LHC are expected more than 280 events 
per year provided that de mass of bileptons is less than 
1 TeV~\cite{Sasaki95}; 2) 
An interesting connection constraining U(1)$_{\rm em}$ electromagnetic gauge 
invariance and the nature of neutrino is realized. In some $G_{3_C3_L1_X}$ 
models the masslessness of the photon prevents the neutrino from acquiring 
Majorana mass~\cite{Ozer94}; 3) 
The $G_{3_C3_L1_X}\rightarrow G_{3_C2_L1_Y}$ breaking scale is estimated by 
running $\sin^2\theta_W$ towards large values which give the upper bound 
of 1.7 TeV~\cite{Ng94} and thus new physics may possibly be in the range 
accessible to accelerator experiments; 4) Lepton number may be explicitly 
broken by trilinear scalar self couplings. This leads to neutrino masses 
proportional to the cube of the correspondig charged lepton mass, with 
consequences for solar neutrinos and for hot dark matter~\cite{Frampton94}; 
4) The $G_{3_C3_L1_X}$ is a phenomenologically viable symmetry for having 
large magnetic moment for the electron neutrino while keeping its mass 
naturally small, needed in one proposed solution for the solar neutrino  
problem~\cite{Barbieri89}; 5) The Yukawa couplings of the $G_{3_C3_L1_X}$ 
models automatically contain a Peccei-Quinn symmetry. This symmetry can be 
extended to the entire Lagrangian solving the strong CP problem~\cite{Pal95}; 
6) Conservation of the leptobarion number $F \equiv  B + L$ forbids the 
existence of massive neutrinos and the neutrinoless double beta decay.   
Explicit or/and spontaneous breaking of $F$ implies that the neutrinos have 
an arbitrary mass. The neutrinoless double beta decay also has channels 
that do not depend explicitly on the neutrino mass~\cite{Pleitez93}; 
7) A supersymmetric $G_{3_C3_L1_X}$ model has a possible reduction to 
the standard $G_{3_C2_L1_Y}$ model with two doublets at the electroweak 
energy scale. Because of the existence of cubic invariants in the 
superpotential of the larger theory, the reduced Higgs potential is not 
that of the minimal supersymmetric standard model~\cite{Duong93}. 
To explore the Higgs sector 
at the electroweak energy scale, it is important to realize that even if 
supersymmetry exists, the minimal supersymmetric standard model is not the 
only possibility for two Higgs doublets. A first example based on E$_6$ 
particle content left-right supersymmetric model has already been 
discovered~\cite{Ma95}; 8) Finally, notice that using the 
lightest leptons as the particles which 
determine the approximate symmetry, if each family of fermions is treated 
separately, the SU(4)$_L$ is the highest symmetry group to be considered 
in the electroweak sector~\cite{Voloshin}. Here we find the Ockam razor 
for direct chiral gauge extensions without exotic charged 
leptons.        
\par
Let us examine the possible cancellation among vacuum contribution terms  
in the  
framework of $G_{3_C3_L1_X}$ gauge extensions. Consider the simplest case     
where the scalar fields are attributed as~\cite{Pisano92,PleTon93}  
\begin{equation}
\eta = \begin{array}{cccccc}
\left(\begin{array}{c}
\eta^0 \\ \eta^-_1 \\ \eta^+_2
\end{array}\right) \sim ({\bf 1},{\bf3},0); \qquad
\rho = \left(\begin{array}{c}
\rho^+ \\ \rho^0 \\ \rho^{++}
\end{array}\right) \sim ({\bf 1},{\bf3},1); \qquad
\chi = \left(\begin{array}{c}
\chi^- \\ \chi^{--} \\ \chi^0\end{array}\right) \sim ({\bf 1},{\bf3},-1),
\end{array}
\label{tri}
\end{equation}
which give the following pattern of symmetry breaking     
\par
\medskip
\centerline{SU(3)$_L\otimes$ U(1)$_N$ 
$\stackrel{\langle\chi\rangle}{\longrightarrow}$ SU(2)$_L\otimes$
U(1)$_Y$ $\stackrel{\langle\rho,\eta\rangle}{\longrightarrow}$ U(1)$_{\rm em}$.}
 
The neutral components of the Higgs triplets of the Eqs. (\ref{tri}) develop the 
vacuum expectation values $v_\eta$, $v_\rho$ and $v_\chi$, respectively, with 
$\left(246 \mbox{ GeV}\right)^2 \equiv v_W^2 = v_\eta^2 + v_\rho^2$. Since
the fields are worked as pertubations around the stable vacuum we define
\begin{equation}
\varphi = v_\varphi + \xi_\varphi + i\zeta_\varphi,
\label{desloc}
\end{equation}
$\varphi = \eta^0$, $\rho^0$, $\chi^0$.     
With the three triplets of the Eqs. (\ref{tri}) we can 
write the more general, renormalizable and gauge invariant Higgs potential
\begin{eqnarray}
V_T(\eta,\rho,\chi)& = &\mu_1^2\eta^\dagger\eta+\mu^2_2\rho^\dagger\rho +
\mu^2_3\chi^\dagger\chi+
\lambda_1(\eta^\dagger\eta)^2 + \lambda_2(\rho^\dagger\rho)^2 + 
\lambda_3(\chi^\dagger\chi)^2 
+\nonumber
\\ & &\mbox{}
+(\eta^\dagger\eta)\left[\lambda_4(\rho^\dagger\rho )+\lambda_5(\chi^\dagger
\chi)\right] +\lambda_6(\rho^\dagger\rho)(\chi^\dagger\chi) + 
\lambda_7(\rho^\dagger\eta)(\eta^\dagger\rho) \nonumber +\\ & &\mbox{} + 
\lambda_8(\chi^\dagger\eta)(\eta^\dagger\chi) + 
\lambda_9(\rho^\dagger\chi)(\chi^\dagger\rho) +
(\frac{f_1}{2}\varepsilon^{ijk}\eta_i\rho_j\chi_k + \mbox{H. c.}),
\label{potential1}
\end{eqnarray} 
where the $\mu$'s, $\lambda$'s and $f_1$ are coupling constants. The 
leptobarionic number $F = L + B$ is conserved in Eq.(\ref{potential1}), where 
$L$ and $B$ are the total leptonic and barionic numbers, respectively
~\cite{Pleitez93}. In 
order to avoid linear terms in $\varphi$ fields the potential shifted 
according to Eq. (\ref{desloc}) requires the validity of the relations  
\begin{mathletters}
\begin{eqnarray}
\mu^2_1 + 2\lambda_1v_\eta^2 + \lambda_4v_\rho^2 + \lambda_5v_\chi^2 + 
f_1\frac{v_\rho v_\chi}{2v_\eta} & = & 0, 
\label{vinc1}
\\
\mu^2_2 + 2\lambda_2v_\rho^2 + \lambda_4v_\eta^2 + \lambda_6v_\chi^2 + 
f_1\frac{v_\eta v_\chi}{2v_\rho} & = & 0, 
\\
\mu^2_3 + 2\lambda_3v_\chi^2 + \lambda_5v_\eta^2 + \lambda_6v_\rho^2 + 
f_1\frac{v_\eta v_\rho}{2v_\chi} & = & 0.
\label{vinc3}
\end{eqnarray}
\label{vinculos1}
\end{mathletters}
where {Im}$f_1 = 0$.     
The scalar sector of the $G_{3_C3_L1_X}$ model which we    
are considering here was carried
 
out in             
Ref.~\cite{Ton96} in the approximation $v_\chi \approx -f_1 \gg v_\eta,
v_\rho$. This approximation leads to the conditions   
\begin{equation}
\lambda_4 \approx 2\frac{\lambda_2v_\rho^2 - \lambda_1v_\eta^2}{v_\eta^2 - 
v_\rho^2}, \qquad 
\lambda_5v_\eta^2 + 2\lambda_6v_\rho^2 \approx -\frac{v_\eta v_\rho}{2}.
\label{condap}
\end{equation}
Since we are interested in the $G_{3_C3_L1_X}$ 
vacuum contribution to the cosmological constant
let us replace Eq. (\ref{desloc}) in Eq. (\ref{potential1}) and use the
constraints (\ref{vinculos1}) and (\ref{condap}). Eliminating terms representing
vacuum flutuations and maitaining only terms in lower order in $1/v_\chi$ we 
obtain, after to require the vanishing of the remaining potential terms,  
\begin{equation}
v_\chi = v_\eta\left[\frac{\lambda_1\left(4v_\eta^2 - 
3v_W^2\right)}{\lambda_3\left(v_W^2 - 
2v_\eta^2\right)}\right]^{1/4} 
\label{vchi}
\end{equation}
 $v_\eta$ and $\lambda_1/\lambda_3$ ratio. 
where $\lambda_3 < 0$ in order to obtain real square masses for the 
neutral scalar bosons~\cite{Ton96}.     
The model of the Refs.~\cite{Pisano92} requires the sextet  
\begin{equation}
S=\left(\begin{array}{ccc}
\sigma^0_1 & s_2^+ & s_1^- \\
s_2^+ & S_1^{++} & \sigma_2^0 \\
s_1^- & \sigma_2^0 & S_2^{--} \\ 
\end{array}\right) \sim \left({\bf 1},{\bf 6}^*, 0\right)
\label{sextet}
\end{equation}
of scalar fields besides the triplets of the Eqs. (\ref{tri}). Here we are 
considering $\langle\sigma_1\rangle = 0$ for maintaining zero neutrino    
mass~\cite{Pleitez93}. Thus, the Eqs. (\ref{desloc})    
have only one more component
$\sigma_2 = v_2 + \xi_2 + i\zeta_2$, where $\langle\sigma_2\rangle = v_2$. 
In this case we get the Higgs potential as
\begin{eqnarray}
V_S\left(\eta , \rho , \chi, S\right) & = & V_T + \mu^2_4
\mbox{Tr}\left(S^\dagger 
S\right) + \lambda_{10}\mbox{Tr}^2\left(S^\dagger S\right) + 
\lambda_{11}\mbox{Tr}\left[\left(S^\dagger S\right)^2\right] + \cr
&& + \left[\lambda_{12}\left(\eta^\dagger\eta\right) + 
\lambda_{13}\left(\rho^\dagger\rho\right) + 
\lambda_{14}\left(\chi^\dagger\chi\right)\right]
\mbox{Tr}\left(S^\dagger S\right) + \cr
&& +(\frac{f_2}{2}\rho_i\chi_jS^{ij} + \mbox{H.c.}),
\label{potential2}
\end{eqnarray}
where $V_T$ is given in Eq.(\ref{potential1}). The constraint equations
equivalent to Eqs. (\ref{vinculos1}) are expressed now by 
\begin{mathletters}
\begin{eqnarray}
\mu_1^2 + 2\lambda_1v_\eta^2 + \lambda_4v_\rho^2 + \lambda_5v_\chi^2 +
              2\lambda_{12}v_2^2 + f_1\frac{v_\eta v_\chi}{2v_\eta} & = & 0,\\
\mu_2^2 + 2\lambda_2v_\rho^2 + \lambda_4v_\eta^2 + \lambda_6v_\chi^2 +
              2\lambda_{13}v_2^2 + f_1\frac{v_\eta v_\chi}{2v_\rho} + 
              f_2\frac{v_\chi v_2}{2v_\rho} & = & 0,\\
\mu_3^2 + 2\lambda_3v_\chi^2 + \lambda_5v_\eta^2 + \lambda_6v_\rho^2 + 
              2\lambda_{14}v_2^2 + f_1\frac{v_\eta v_\rho}{2v_\chi} + 
              f_2\frac{v_\rho v_2}{2v_\chi} & = & 0,\\
\mu_4^2 + 2(2\lambda_{10} + \lambda_{11})v_2^2 + \lambda_{12}v_\eta^2 + 
             \lambda_{13}v_\rho^2 + \lambda_{14}v_\chi^2 +
             f_2\frac{v_\rho v_\chi}{4v_2} & = & 0.
\end{eqnarray}
\label{vinculos2}
\end{mathletters}
The approximation
\begin{equation}
v_\chi \approx -f_1 \approx \vert f_2\vert \gg v_\eta, v_\rho, 
v_2, 
\label{ap}
\end{equation}
with $v_W^2 = v_\eta^2 + v_\rho^2 + v_2^2$ leads to~\cite{Ton96}
\begin{equation}
4\left(2\lambda_{10} + \lambda_{11}\right) \approx \lambda_1 
\approx \lambda_2 \approx \lambda_{12} \approx \lambda_{13} \approx 
\frac{\lambda_4}{2}.
\label{condap2}
\end{equation}
Following analogous procedure as in the three triplets case we replace Eqs. 
(\ref{ap}) and (\ref{condap2}) in the potential $V_S$ in Eq. (\ref{potential2})
and maitain only the vacuum expectation value contributions.     
Next, if we require that the 
remaining  
potential terms vanish we obtain
\begin{equation}
\left(-\lambda_6 + \lambda_9\right)v_W^2 + \left(-1 - \lambda_5 + \lambda_6 + 
\lambda_8\right)v_\eta^2 - \lambda_3v_\chi^2 +   
\left(-3\lambda_{14} + \lambda_6 - \lambda_9\right)v_2^2 = 0. 
\label{vs0}
\end{equation}
This is the condition for an arbitrarily small cosmological constant in 
the case where the symmetric sextet of scalar fields play an essential role 
in order to generate the charged lepton masses. 

\par
We have presented a way for bringing the energy density of the vacuum to zero,  
associated with the flat empty world of Minkowski, in a gauge model of 
elementary particles. In all elementary particle theories, including 
supersymmetry and grand unification, it is predicted a very large value.   
Specially, we find remarkable that a vanishing    
cosmological constant can be obtained in a relatively simple extension        
of the standard model which has several replys for some fundamental    
questions that plague the physics of elementary particles. However, 
even if it is shown that the value of $\Lambda$ can be zero, the 
cosmological constant problem still remains, whether $\Lambda$ actually 
does vanish exactly and identically due to some symmetry or another 
unknown deep physical connection.

\acknowledgments
This work was supported by the Funda\c c\~ao de Amparo \`a Pesquisa do Estado 
de S\~ao Paulo (FAPESP), the Instituto de F\'\i sica Te\'orica (IFT), and by 
the Universidade do Estado do Rio de Janeiro (UERJ).

\end{document}